\pdfoutput=1

\documentclass[11pt]{article}

\usepackage{acl}

\usepackage{times}
\usepackage{latexsym}
\usepackage{graphicx}
\usepackage{newfloat}
\usepackage{listings}
\usepackage{amsmath}
\usepackage{amsfonts}
\usepackage{multirow}
\usepackage{booktabs}
\usepackage{amssymb}
\usepackage{algorithm, algpseudocode}
\usepackage{soul}
\usepackage[T1]{fontenc}

\usepackage[utf8]{inputenc}

\usepackage{microtype}

\usepackage{inconsolata}

\title{RePair: Automated Program Repair with Process-based Feedback}

\author{Yuze Zhao$^1$,  Zhenya Huang$^{1,2}$\thanks{Corresponding author},  Yixiao Ma$^1$,  Rui Li$^1$,  Kai Zhang$^1$,\\ {\bf  Hao Jiang$^1$,  Qi Liu$^{1,2}$, Linbo Zhu$^{1,2}, $ Yu Su$^{2,3}$}\\
$^1$State Key Laboratory of Cognitive Intelligence, University of Science and Technology of China \\ $^2$Institute of Artificial Intelligence Comprehensive National Science Center\\
$^3$School of Computer Science and Artificial Intelligence, Hefei Normal University, China\\
\small \texttt{\{yuzezhao, mayx, ruili2000, kkzhang0808, jianghao0728\}@mail.ustc.edu.cn} \\ \small \texttt{\{huangzhy, kkzhang08, qiliuql\}@ustc.edu.cn} \\ \small \texttt{lbzhu@iai.ustc.edu.cn} \\ \small \texttt{yusu@hfnu.edu.cn}
}

\begin{document}
\maketitle
\begin{abstract}
The gap between the trepidation of program reliability and the expense of repairs underscores the indispensability of Automated Program Repair (APR).
APR is instrumental in transforming vulnerable programs into more robust ones, bolstering program reliability while simultaneously diminishing the financial burden of manual repairs.
Commercial-scale language models (LM) have taken APR to unprecedented levels.
However, the emergence reveals that for models fewer than 100B parameters, making single-step modifications may be difficult to achieve the desired effect.
Moreover, humans interact with the LM through explicit prompts, which hinders the LM from receiving feedback from compiler and test cases to automatically optimize its repair policies.
In this literature, we explore how small-scale LM (less than 20B) achieve excellent performance through process supervision and feedback.
We start by constructing a dataset named CodeNet4Repair, replete with multiple repair records, which supervises the fine-tuning of a foundational model.
Building upon the encouraging outcomes of reinforcement learning, we develop a reward model that serves as a critic, providing feedback for the fine-tuned LM's action, progressively optimizing its policy.
During inference, we require the LM to generate solutions iteratively until the repair effect no longer improves or hits the maximum step limit.
The results show that process-based not only outperforms larger outcome-based generation methods, but also nearly matches the performance of closed-source commercial large-scale LMs. \footnote{Code and data are publicly available at \url{https://github.com/TnTWoW/RePair}}
\end{abstract}

\section{Introduction}
Since the birth of the program, its reliability has been a primary concern. The capacity of large language models to auto-generate code has further intensified these  concerns~\cite{khoury2023secure}, sparking discussions on possible solutions.
Program repair accepts vulnerable programs as input, enhancing them through locating, correcting, and testing - a process that is often challenging within software development and programming competition.
The confrontation arises when these defective programs encounter errors only when they receive rare and unexpected inputs. Moreover, it requires experienced programmers to invest substantial time in identifying and rectifying errors.
A sophisticated \textbf{A}utomated \textbf{P}rogram \textbf{R}epair (APR) system can reduce the expertise threshold and time commitment required, significantly enhancing the robustness of the program.

\begin{figure}[t]
    \centering
    \includegraphics[width=\columnwidth]{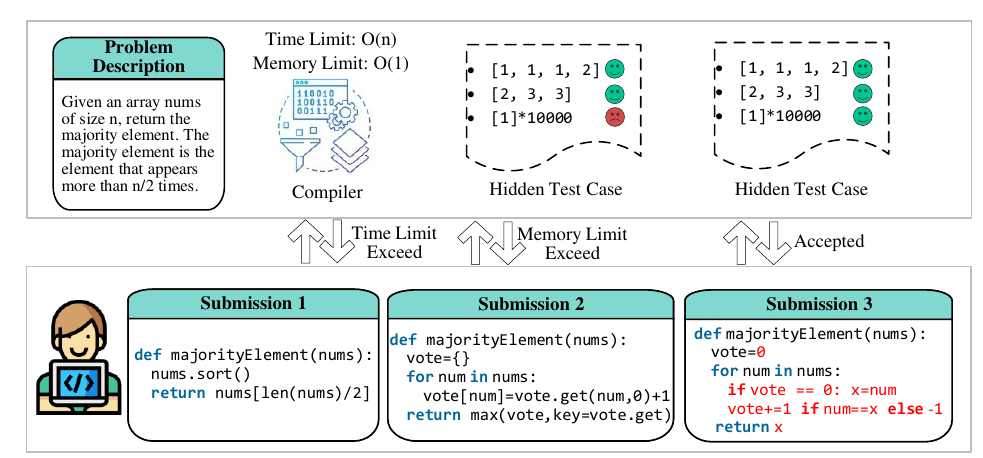}
    \caption{The general procedure for competitors to refine a solution on programming contest platform. Initially, they draft a solution based on the problem description and additional constraints such as time and memory limits. They then progressively improve their solution using feedback from the platform, like exceeding time or memory limits, until they achieve an accepted result.}
    \label{multi-turn repair example}
\end{figure}

Prior works viewed APR as a sequence-to-sequence task, then LLMs, with their massive model parameters and training data, can address this issue through zero- or few-shot learning~\cite{Jiang2021cure, openai2023gpt4, ouyang2022training, chen2021evaluating, zeng2022glm, touvron2023llama, hoffmann2022training, anil2023palm, chen2022codet, le2022coderl, rozière2024code, guo2024deepseekcoder, wei2023copiloting, zhang2023gamma}.
Whether it's traditional methods or LLMs, they achieve the repair by making a single modification.
These outcome-based supervision methods pose a stringent challenge to the model due to the substantial edit distance between input and output. 
Simultaneously, the one-step modification approach does not align with human behavioral patterns: To be specific, when facing complex tasks, programmers usually repair the buggy program step by step in a cycle of modification, testing, and feedback.
This inspires us to pay attention to the exploration of modification processes in APR tasks. Unlike \textit{outcome-based} supervision, this \textit{process-based} method requires guidance and supervision at each modification step.

Figure~\ref{multi-turn repair example} depicts the typical program repair process in programming competitions, which is guided by compilers and test cases: Competitors initially construct a sketch of the program based on the problem's intent and other constraints. They then complete the repair through a continuous process of submission, feedback, and interaction. 
Specifically, they initially attempt to solve ``Find the majority element'' problem by sorting and selecting the middle element.
However, when they encounter time constraints, they shift to calculating each element's frequency. But this method still requires $O(n)$ space complexity.
After receiving ``Memory Limit Exceed'' status, they ultimately use the Boyer-Moore majority vote algorithm with $O(1)$ space complexity.

This example illustrates the characteristics of \textit{process-based} program repair, which necessitates ongoing interaction with the compiler and test cases for feedback-guided repair.
However, applying this type of process-based feedback on LMs appears unfeasible.
First and foremost, the primary obstacle hindering related research is the absence of process-based datasets in practical scenarios.
Second, the supervision method for intermediate processes in APR tasks are still under exploration.
Previous work explored process-based supervision in mathematical problem solving and reasoning tasks~\cite{lightman2023lets, liu2023guiding}, which required the intermediate steps to be correct. However, the intermediate steps in APR tasks serve as incorrect supervision signals and cannot be directly used for training.
Last, interaction with LMs can be achieved through explicit prompt engineering. 
Attempts to use prompt engineering instead of compiler and test cases for feedback hinder the precise construction of human intentions by LMs.

In this work, to the best of our knowledge, we conduct the first few comprehensive exploration of process-based feedback with LMs in APR task.
For this, we first establish a multi-step program repair dataset called CodeNet4Repair. 
Following that, we introduce a process-based feedback APR framework called \textbf{RePair}.
RePair includes two models: a reward model and a repair model.
We start by training a reward model to mimic the compiler as a virtual tool. It takes program text as input and gives assertions about the program's status.
The repair model works on buggy programs, completing one repair in a step. It then offloads the assessment of the program's status to the virtual tool and waits for feedback to adjust the strategy for the next modification.
Finally, we use pass@$k$ as a metric to objectively assess the quality of program repair.

\begin{table}[t]
\setlength\tabcolsep{3pt}
    \centering
   \resizebox{\linewidth}{!}{
   \begin{tabular}{lcccc}
                 & DeepFix  & Review4Repair & Bug2Fix  & CodeNet4Repair \\ \hline
Language  & C & Java& Java& Python         \\
Test Cases   & $\times$ & $\times$& $\times$& \checkmark            \\
Repair Form  & single step & single step & single step & multi step     \\
Problem Description & $\times$ & $\times$ & $\times$& \checkmark            \\
Test Size   & 6,971& 2,961 & 58,356,545 & 10,144          \\ 
\hline
\end{tabular}}
    \caption{A comparison between the CodeNet4Repair dataset and existing datasets for program repair. The advantages of CodeNet4Repair stem from its comprehensive 
 inclusion of test cases, problem descriptions, and detailed repair steps. CodeNet4Repair's test set contains 10,144 complex program repair processes at competition level. }
    \label{tab:dataset_compare}
\end{table}

\section{Data Collection}
As far as we know, there has not been a dataset established for program repair tasks that includes processes. Thus we attempt to develop a procedural dataset specifically for these tasks.
Overall, the dataset includes problem description, memory and time limit of the problem, repair process (a series of programs from error to correctness) and resource usage during execution (memory usage, CPU time, and code size).
We derive our data from the large-scale programming competition dataset CodeNet~\cite{puri2021project}.
We make every effort to cleanse and filter this data to guarantee its quality.
We clarify problem description, clean up the program from preliminary to fine, and collect additional high-quality test cases. Finally, we organize the program into a procedural format.
In Table~\ref{tab:dataset_compare}, we compare other excellent datasets and showcased the unique aspects of CodeNet4Repair~\cite{Rahul2017DeepFix, huq2020review4repair, tufano2019learning}.
\begin{figure*}
    \centering
    \includegraphics[width=0.9\textwidth]{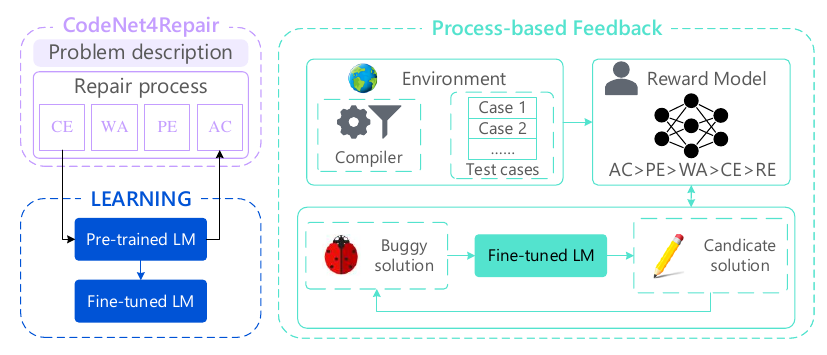}
    \caption{An illustration of process-based Automated Program Repair with compiler and test case feedback: (1) The introduction of a clean, privacy-protected dataset called CodeNet4Repair. (2) The application of Supervised Fine-Tuning (SFT) on pre-trained language models. (3) The incorporation of process-based feedback via reinforcement learning (RL). This process includes: establishing a reward model as a critic, and LM adjusts its repair policies based on the feedback from the critic. SFT and RL are both trained on CodeNet4Repair training set.}
    \label{model}
\end{figure*}
\subsection{Problem Description Collection}
Even professional programmers need to understand the problem that the program is intended to solve when fixing it.
CodeNet provides unprocessed HTML for storing the description of each question.
We utilize regular matching to extract these descriptions from the CodeNet's HTML files.
Finally, we manually supplement any descriptions that were challenging to recognize with Optical Character Recognition.

\subsection{Program Preliminary Filtering}
We chose Python as our benchmark language.
We design the following rules for preliminary filtering.

(1) No duplicate submissions. Casual or copy submissions do not aid in program repair process.

(2) No malicious submissions, which include automated disruptive attempts, invalid code aimed at attacking the platform's codebase, and destructive codes designed to interfere with system files.

(3) No privacy breach submissions. Some IDEs automatically generate comments that could reveal the author's identity.

(4) Ensure each repair process includes at least one acceptable commit. The processes without any accepted outcomes will be excluded.

After this preliminary filtering, we obtain 1,227,259 submission programs. All non-English characters from the code comments are removed prior to fine filtering.

\subsection{Program Fine Filtering}
Although the original CodeNet dataset provides the program execution status (e.g., Wrong Answer or Accepted), these may vary due to differences in Python versions and environments.
We set up a standard Python 3.11.3 environment and are meticulously testing each of the 1,227,259 execution results using test cases to ensure consistent status. 
Ignoring extra expenses such as process switching and assuming an average execution time of 4 seconds per solution, we would require a total of $\frac{4 \times 1227259}{60 \times 60}=1364 (Core\cdot hours)$ CPU time.
Finally, we obtain 278,408 consistent programs. 
We arrange the answer records of each user for each question in chronological order, ensuring that the last program is ``Accepted''.

To enhance evaluations precision, we expand the number of high-quality test cases collected online and annotated by hand. We organize these programs with high-quality test cases into a procedure-based dataset called CodeNet4Repair. CodeNet4Repair is an information-leak-free program repair dataset with Apache-2.0 License.

\section{Method}
Our methodology follows that of \citeauthor{ouyang2022training}, which aligned GPT-3 with human.
We start from StarCoderBase~\cite{li2023starcoder}, a 15.5B code foundation pre-trained LM with 8K context length. It is trained on 1 trillion tokens from The Stack~\cite{kocetkov2022stack}.
Figure~\ref{model} illustrates the techniques used to train RePair.
\subsection{Supervised Fine Tuning on APR Task}
To ensure that the LM can understand program repair tasks, we use the prompt templates in Appendix~\ref{sec:template} for Supervised Fine-Tuning (SFT)~\cite{zhang2022incorporating, huang2024edunlp}.
Given a vulnerable program sequence $\mathbf{x}=\{ x_1,x_2,...,x_N\}$, LM is expected to output a robust program $\mathbf{y}=\{ y_1,y_2,...,y_{N^{'}}\}, y_t \in \mathcal{V}$ that can pass all test cases, where $\mathcal{V}$ is vocabulary. During the training phase, the model parameters $\theta$ are learned by maximizing the likelihood of the output and the ground-truth~\cite{zhang2021eatn}. The training objective is to minimize the following loss:
\begin{equation}
\begin{aligned}
\mathcal{L}_{ce}(\theta)&=\mathbb{E}[-\log P(\mathbf{y} \mid \mathbf{x} ; \theta)] \\
&=-\sum_{t=1}^{N^{\prime}} \log P\left(\mathbf{y}_{t} \mid \mathbf{y}_{<t}, \mathbf{x} ; \theta\right),
\end{aligned}
\end{equation}
where $\mathbb{E}$ is the expectation over entire dataset, and $\mathbf{y}_{ < t}$ is a partial sequence before time-step $t$.
\subsection{Process-based Feedback}
After supervised fine-tuning, we generally obtain a fine-tuned LM suitable for the APR tasks.
However, this model only understands how to provide a possible solution under the condition of given $\mathbf{x}$, lacking process supervision and feedback from environment like compilers and test cases.
To ensure that the LM can gradually refine the program through interaction, we introduce reinforcement learning (RL).
In this context, we treat the fine-tuned LM as an \textit{actor}. Given a \textit{state} $\mathbf{x}$, its output $\mathbf{\hat{y}}$ is considered an \textit{action}. Guided by feedback from the reward model, it iteratively refines program towards a final possible result.
\subsubsection{Reward Modeling}
Instead of using direct feedback from the compiler and test cases, we train a reward model (RM) to assess program quality.
The reward model serves as a \textit{virtual tool}, while the LM optimizes repair strategies by interacting with it.
The main reasons boil down to two points: (1) During training, direct interaction with the environment significantly blocks the training pipeline. 
The batch-processed tensors in cuda must first be transferred to memory for decoding, undergoing syntax check and case testing sequentially; when providing feedback, these results are re-encoded and sent back to cuda. This process drastically reduces the throughput - a cost that is unbearable.
(2) During the inference phase, the inability to access test cases for vulnerable programs compromises generalization performance.

Unlike the methods used by (\citeauthor{pail2017deep, ziegler2020finetuning, stiennon2022learning, ouyang2022training}), our approach doesn't require extensive investment in collecting human preferences to train a reward model. We can substantially reduce costs by sourcing program execution preferences from automated compiler and test cases. 
Specifically, we first empirically define a non-strict partial order based on program quality from high to low:
\begin{equation}
\nonumber
\text{AC}>\text{PE}>\text{WA}=\text{TLE}=\text{MLE}>\text{CE}>\text{RE}.
\end{equation}

AC signifies that the program was ``Accepted'' by all test cases. PE stands for ``Presentation Error", which indicates that the output data is correct but not properly formatted.
WA, TLE, MLE, CE, RE are abbreviations for ``Wrong Answer'', ``Time Limit Exceed'', ``Memory Limit Exceed'', ``Compile Error'' and ``Runtime Error'' respectively.
This partial order defines the severity of program errors in an ascending order, from AC (lowest) to RE (highest).
We prioritize fixing RE over CE because RE typically indicates more serious issues such as logical errors which pose a greater risk than CE.

\begin{algorithm}[t]
\textbf{Input}: Trained critic $r_\phi(\cdot)$; Trained actor's policy $\pi_{\theta}$; Vulnerable program $x_0$; Max iterations $T$;  Max Patience $P$\\
\textbf{Output}: Repaired Program
\begin{algorithmic}[1] 
\State $t=0$ \Comment{Counter of timestep}
\State $p=0$ \Comment{Counter of unimproved patience}
\State $r_0=r_\phi(x_0)$ \Comment{Initialize the reward}
\While{$t<T$ $\And$ $p<P$}
\State $x_{t+1} = \pi_{\theta}(x_t)$ \Comment{Actor generate}
\State $r_{t+1} = r_\phi(x_{t+1})$ \Comment{Critic review}
\State $\Delta=r_{t+1}-r_t$ \Comment{Evaluate the boost}
\If{$\Delta \leq 0$ } \Comment{If no boost}
\State $p=p+1$
\Else
\State $p=0$ \Comment{Counter reset}
\EndIf
\State $t=t+1$
\EndWhile
\State \textbf{return} $x_{t-p}$ \Comment{Roll back to $p$ steps ago}
\end{algorithmic}
\caption{Generate Repaired Program with Process-based Feedback, Actor-Critic Style}
\label{process feedback}
\end{algorithm}

Then, we optimize the reward model parameters $\phi$ using pairwise ranking based on the above partial order.
Given $K$  programs with status, all aimed at solving an identical problem, there is a partial order set $D$ containing $\tbinom{K}{2}$ pairs of partial orders. After shuffling $D$, we train the reward model by minimizing the following pairwise ranking loss:
\begin{equation}
\resizebox{0.48\textwidth}{!}{$
    \mathcal{L}_{pr}(\phi)=-\frac{1}{\binom{K}{2}} \mathbb{E}_{\left(y_w, y_l\right) \sim D}\left[\log \left(\sigma\left(r_\phi\left(y_w\right)-r_\phi\left(y_l\right)\right)\right)\right],$}
\end{equation}
where $r_\phi\left(\cdot\right)$ is a reward model, which takes a program as input and outputs a scalar reward.
In practice, we use another smaller pre-trained auto-regressive model, and replace the original model's non-embedding layers with a projection layer to output a scalar value.
The pair $\left(y_w, y_l\right)$ is a partial order where $y_w$ is the preferred program (for instance, the status of $y_w$ is AC while $y_l$ is WA). 

\begin{figure}[t]
    \centering
    \includegraphics[width=0.48\textwidth]{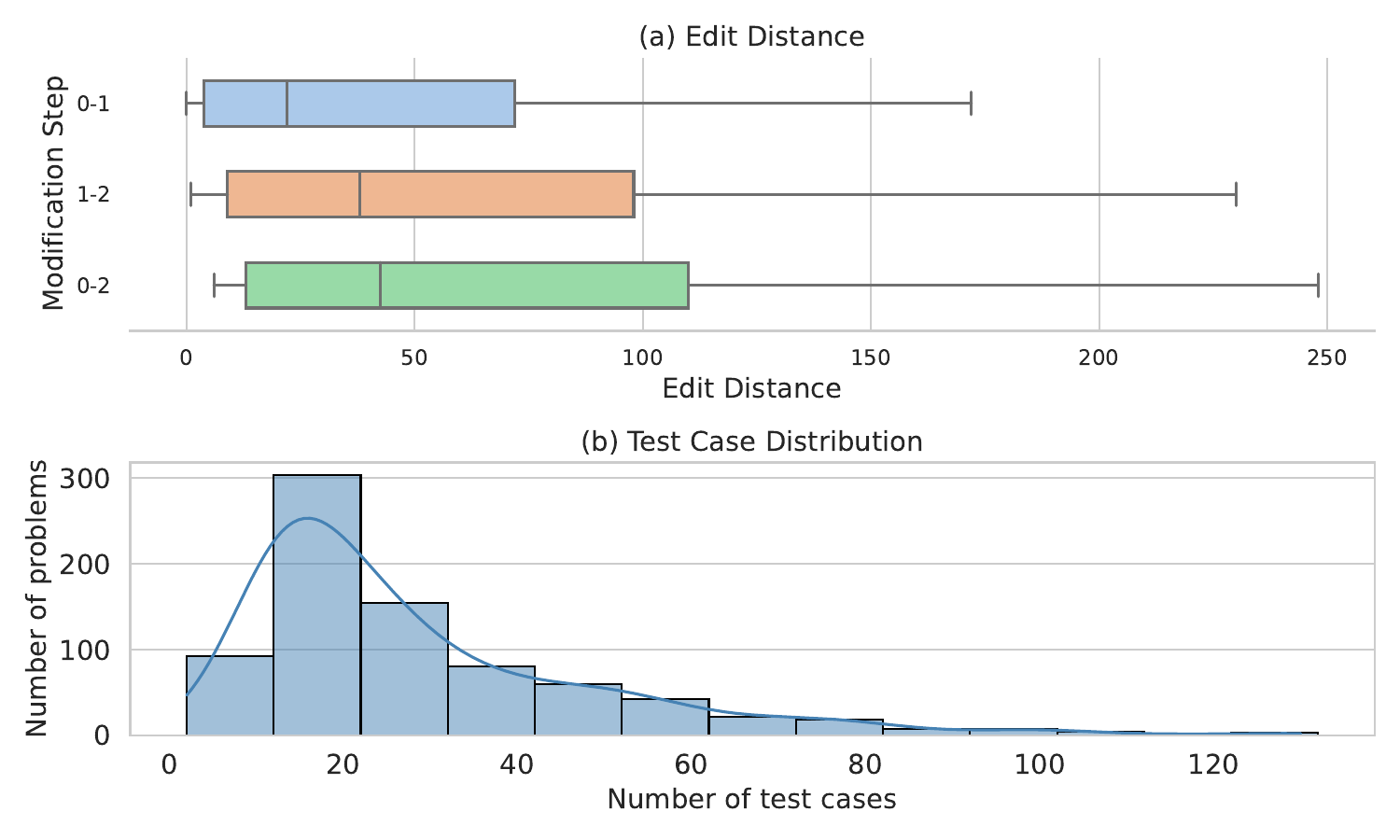}
    \caption{(a) The edit distances between three program after 2-step modifications. 0-1: The edit distance after the first refinement; 1-2: The edit distance after the second refinement; 0-2: The edit distance of a single-step refinement. (b) The distribution of test cases. Most of the test cases are concentrated between 10-20.}
    \label{test cases distribution}
\end{figure}

\subsubsection{Reinforcement Learning}
In the previous two sections, we fine-tuned a LM and developed a reward model capable of realistically evaluating program states by optimizing loss $\mathcal{L}_{ce}(\theta)$ and loss $\mathcal{L}_{pr}(\phi)$ respectively.
In this section, we will utilize reinforcement learning (RL) to finalize program repair through multiple interactions with the fine-tuned LM and well-trained reward model~\cite{qian2021derivative}.
We adopt Proximal Policy Optimization (PPO) algorithm~\cite{stiennon2022learning, ouyang2022training, schulman2017proximal} to fine-tune our LM. 
The LM functions as an actor, generating a repaired program (\textit{action}) based on the input program (\textit{state}) and the \textit{policy} $\pi$ of the LM (i.e., LM parameters $\theta$). The trained reward model plays as a critic, assessing the effectiveness of repairs and rewarding the LM accordingly. Our objective is to optimize the LM parameters $\theta$ by maximizing these rewards.

To be more specific, we supervise the process by maximizing the the following objective function:
\begin{equation}
\resizebox{0.48\textwidth}{!}{$\begin{aligned}
&\mathcal{L}_{rl}(\theta)\!=\!\mathbb{E}_{\left(x_i, x_{i+1}\right) \sim D_{\pi_{\theta^{'}}}}\!\left[r_\phi(x_{i+1})\!-\!r_\phi(x_{i})\!-\!\beta \text{KL}\left(\theta,\theta^{'}\right)\right]\\
&=\mathbb{E}_{\left(x_i, x_{i+1}\right) \sim D_{\pi_{\theta^{'}}}}\left[\Delta^r_{\phi}(x_{i+1}, x_{i})-\beta \log \frac{\pi_{\theta}(x_{i+1} \mid x_{i})}{\pi_{\theta^{'}}(x_{i+1} \mid x_{i})}\right],
\end{aligned}$}
\label{loss:rl}
\end{equation}
where $x_{i+1}$ is the refined output based on the input $x_i$ and learned LM's policy $\pi_{\theta^{'}}$, $r_\phi(\cdot)$ is calculated as the learned reward model, the KL reward coefficient $\beta$ control the strength of the KL penalty, $\Delta^r_{\phi}(x_{i+1}, x_{i})=r_\phi(x_{i+1})-r_\phi(x_{i})$ means the evaluation of the degree of improvement. 

\subsection{Multi-step Generation Under RM Supervision}
We achieved process supervision with feedback by maximizing reward function. In this section, we will demonstrate how the language model interacts with virtual tools during the generation phase to achieve alignment during both training and generation stages. We continuously request the LM to refine solutions until either of two conditions is met: (1) there is no improvement ($\Delta\leq0$) in $P$ consecutive steps; (2) the maximum number of iterations has been reached. The Algorithm~\ref{process feedback} elucidates the specific process of generation.
\section{Experiments}
\subsection{Data Preparation} 
We sample repair records that corrected after two-step modifications from CodeNet4Repair, resulting in three versions, and calculate the edit distance between these versions. In Figure~\ref{test cases distribution} (a), the findings align with our hypothesis: 
The edit distance required to complete the repair in one step is longer than that required in multiple steps.
Despite an increase in overall cost, step-by-step repairing can reduce the complexity of tasks.

\begin{table}[t]
    \centering
    \resizebox{\linewidth}{!}{
\begin{tabular}{cccc}
\hline
\multicolumn{1}{l}{}           & \#problems          & \#records       & \#codes       \\ \hline
\multicolumn{1}{c|}{Train set} & 563                & 94062           & 252031        \\
\multicolumn{1}{c|}{Test set}  & 61                 & 10144           & 26377         \\ \hline
                               & record per problem & code per record & size per code \\ \hline
\multicolumn{1}{c|}{Train set} & 167.07             & 2.68            & 270.42        \\
\multicolumn{1}{c|}{Test set}  & 166.3              & 2.60             & 246.88        \\ \hline
\end{tabular}}
    \caption{Dataset Information}
    \label{tab:dataset_static}
    \vspace{-0.2cm}
\end{table}
\begin{table*}[t]
\setlength\tabcolsep{4pt}
\centering
\resizebox{\linewidth}{!}{
\begin{tabular}{ccc|cccc|cccc|cccc}
\hline
\multicolumn{2}{c}{\multirow{2}{*}{Model}}                          & \multirow{2}{*}{Size} & \multicolumn{4}{c|}{pass@$1$} & \multicolumn{4}{c|}{pass@$3$} & \multicolumn{4}{c}{pass@$5$} \\ \cline{4-15} 
\multicolumn{2}{c}{}                                                &                       & Easy  & Medium & Hard & All & Easy  & Medium & Hard & All & Easy & Medium & Hard & All \\ \hline
\multicolumn{1}{c|}{\multirow{4}{*}{Close-source}} & PaLM & 540B  &  16.83 & 15.94 & 18.57  &  17.13   & 22.07  &  29.38  &   27.86   &  26.11   &   24.39   & 37.50  &   31.43   &  30.56   \\
\multicolumn{1}{c|}{} & GPT-3.5       & -  &    53.41   &    45.32    &  39.14    & 46.39    &   65.24    &   62.03     &   58.57   &   62.13  &   68.29   &   65.63     &   65.71   &  66.67   \\
\multicolumn{1}{c|}{}& Claude2 & - & 50.73  &   43.75     &   36.29   & 43.98 &     65.12  &    63.75    &    55.29  &  61.53   &   71.95   &  73.44  &   62.86   & 69.44    \\
\multicolumn{1}{c|}{}  & ChatGLM-Pro   & -    &    20.49   &  20.94      &   17.43   &   19.63  &   23.17    &     27.19   &    24.43  &  24.77   &  24.39    &     29.69   &   28.57   & 27.31    \\ \hline
\multicolumn{1}{c|}{\multirow{7}{*}{Open-source}}  & StarCoderBase & 15B & 0.49  & 0.31  & 0.00 &   0.28  &   1.46    &   0.94  &  0.00 &  0.83  &   2.44  &  1.56  &   0.00 &  1.39 \\
\multicolumn{1}{c|}{} & StarCoderChat & 15B  & 0.00 & 0.00 & 0.57 &  0.46   &   0.00    &    0.00    &  1.29    &  0.42   &   0.00   &   0.00 &  1.43    &   0.46  \\
\multicolumn{1}{c|}{} & CodeGen2      & 16B   &  4.15     &    1.87    &   0.29   &  2.22   &    10.73   &    5.16    &  0.86    &  5.88   &   15.85   &    7.81    &   1.43   &  8.80   \\
\multicolumn{1}{c|}{} & CodeGeeX2     & 6B    &    7.80   &    7.19    &   1.43   &   5.56  &    19.27   &    17.03    &   3.86   & 13.61    &  26.83    &    23.44    &   5.71   &  18.98   \\
\multicolumn{1}{c|}{} & LLaMA2   & 70B   &  10.24     &    5.94    &  6.00    &  7.59   &  20.37     &    14.84    &  10.86    &   15.65  &   24.39   &    21.88    &    12.86  &   19.91  \\
\multicolumn{1}{c|}{} & LLaMA2-Chat   & 70B   &  30.73     &    23.13   &  18.86   &  24.63   &  39.02   &   33.13   &  25.71   &   32.96  &   41.46   &   39.06   &   30.00 &   37.04  \\
\multicolumn{1}{c|}{}     & \textbf{Our Model }    & 15B    &   \textbf{51.61}    &    \textbf{44.13}    &  \textbf{40.57}   &   \textbf{44.34}   &    \textbf{62.47}  &    \textbf{59.98}    &  \textbf{52.34}    &  \textbf{60.01}  &  \textbf{67.75}    &   \textbf{64.14}     &  \textbf{ 60.32}   &    \textbf{65.66} \\ \hline
\end{tabular}}
    \caption{Results on CodeNet4Repair: compared to open-source models, our process-based feedback method has demonstrated superior performance. It also remains competitive when compared with commercial LLMs.}
    \label{tab:main_result}
\end{table*}
We ensure evaluation quality by using only questions with high-quality test cases for our training and testing sets. We manually annotate and collect additional high-quality test cases from the internet to serve as hidden tests for each problem. This process resulted in test cases for 794 questions.

Figure~\ref{test cases distribution} (b) shows the distribution of these test cases.
To prevent data leakage, we divided CodeNet4Repair based on the problem ID in a ratio of 9:1. The training set consists of 94,062 repair processes, while the testing set comprises 10,144 repair processes.
Table~\ref{tab:dataset_static} provides the statistical information about filtered dataset.
\subsection{Experimental Setup} In the fine-tuning stage, we train our model using mixed precision training. We use AdamW optimizer with 2e-5 learning rate. To prevent training instability caused by an excessively high learning rate, we utilize a cosine LR schedule down to 10\% of the original learning rate with learning rate warmup. We use ZeRO++ to distribute model tensors across accelerators~\cite{wang2023zero}.

In reward modeling, we adopt the LR of 9.6e-6 and cosine learning rate schedule. We randomly select $K=9$ programs in same problem and rank them based on their execution status. Each batch contains 64 units, thus a single gradient backpropagation involves $64 \times \tbinom{9}{2}=2304$ comparisons.

In the process-based feedback stage, we initialize RL policies from the fine-tuned LM. We train the LM for 32k episodes with 512 batch size. We assign a KL penalty factor, $\beta$, of 0.02 and establish a learning rate of 9e-6. 
We perform nucleus sampling with top\_p=0.95 and top\_k=50, and set the temperature to 0.2.

\subsection{Baselines}
We compare the latest state-of-the-art LM that possess few-shot learning and code comprehension capabilities. These models include PaLM (chat-bison-001)~\cite{anil2023palm}, GPT-3.5 (gpt-3.5-turbo-0613), Claude2 and ChatGLM-Pro~\cite{zeng2022glm}. We also compare some open-sourced models such as StarCoderBase/Chat~\cite{li2023starcoder}, CodeGen2, CodeGeeX2~\cite{zheng2023codegeex}, LLaMA2/-chat~\cite{touvron2023llama}. 
\subsection{Evaluation} Previous program repair datasets lacked annotated test cases and stable test environment~\cite{Dinella2020Hoppity, hendrycks2021measuring, li2022competition, wang2023executionbased}, leading to a shortage of automatic evaluation methods for program repair based on execution results. 
We contend this issue and advocate for the evaluation using execution outcomes.

Following previous code generation's evaluation method, we use pass@$k$ as our evaluation metric~\cite{hendrycks2021measuring,chen2021evaluating, nijkamp2023codegen}. Here we use the unbiased estimator proposed in~\cite{chen2021evaluating}, which is defined as follows:
\begin{equation}
\text { pass @ } k=\mathbb{E}\left[1-\frac{\tbinom{n-c}{k}}{\tbinom{n}{k}}\right].
\end{equation}

In this work, we report $k \in \{1,3,5\}$ as the final result. The reason is that in pass@$k$, a larger $k$ provides a more comprehensive assessment of LM's ability to generate code. However, in practice, users won't request multiple candidates repeatedly; therefore, reporting a small number of $k$ suffices for program repair evaluation.
\subsection{Main Results}
We use the pass rate of all problems in 14 billion submission records as a standard to measure the difficulty of the problems. 
These problems are categorized into three levels - easy, medium, and hard - based on their pass rates divided into tertiles.
The model's performance at various difficulties is detailed in Table~\ref{tab:main_result}.
The results of the experiment indicate that our model outperforms all other open-source models. 
Even when compared to sophisticated commercial LLMs, we achieve competitive results. Starting with an analysis of the open-source models:
We acknowledge that some performance gains come from supervised fine-tuning. However, as evidenced by StarCoderBase/Chat results, our foundation model lacks zero- or few-shot program correction capabilities and struggles with repair tasks without supervised fine-tuning.
In comparison to LLMs designed for code downstream tasks, our repair model generates programs that better meet task requirements and outperforms them at every difficulty level.
Moreover, in LLaMA2-Chat we found that multiple rounds of dialogue-based supervised fine-tuning do not degrade but rather enhance its understanding of human instructions and repair abilities.
While our model's performance is not yet on par with commercial LLMs such as ChatGPT and Claude2, considering the resources they've invested - including compute, data collection, and manual annotation among others - which are significantly greater than ours, we have still achieved competitive results.
This validates the effectiveness of our process-based feedback method and provides guidance for future research.
\subsection{Model Analysis}
\begin{table}[t]
    \centering
\setlength\tabcolsep{4pt}
 \resizebox{\linewidth}{!}{
\begin{tabular}{ccc|c|c|c}
\hline
\multicolumn{3}{c|}{Model}        & \multicolumn{1}{c|}{\multirow{2}{*}{pass@1}} & \multicolumn{1}{c|}{\multirow{2}{*}{pass@3}} & \multicolumn{1}{c}{\multirow{2}{*}{pass@5}} \\ \cline{1-3}
Process & Feedback & Reward & \multicolumn{1}{c|}{}                        & \multicolumn{1}{c|}{}                        & \multicolumn{1}{c}{}                        \\ \hline
$\times$           & $\times$      & -      &    36.32  &   47.63    &  52.63      \\
$\times$           & \checkmark      & Pair   &  37.26   &  54.81 &  62.27   \\
\checkmark           & $\times$      & -      &   20.43  &  35.14 &   44.72  \\ \cline{1-3}
\checkmark           & \checkmark      & Pair   & \textbf{44.34}   &  \textbf{60.01} &    \textbf{65.66}   \\ \cline{1-3}
\checkmark           & \checkmark      & Point  & 40.11  &  58.89   &  64.27 \\
\checkmark           & \checkmark      & List  & 39.45  &  57.33 &  63.56 \\
 \hline             
\end{tabular}}
    \caption{The results of ablation study. We conduct an in-depth study on the design of process supervision, feedback, and reward functions. The experimental results confirm our model's effectiveness. Point: Point-wise loss; Pair: Pair-wise loss; List: List-wise loss.}
    \label{tab:ablation}
\end{table}
We explore the rationality of the model including process supervision, feedback, and reward function design.
We design different variants from those three perspectives.
First, we arrange and combine process supervision and feedback to create four different approaches. Then, we discuss the design of reward functions within the process-based feedback framework.

\textbf{Impacts of Process Supervision.} \quad We attempt to eliminate process supervision signals, enabling the model to depend solely on single-step supervision signals. Without feedback, the model degenerates into a fine-tuned single-step program repair model. While with the feedback, the fine-tuned model's adjustment of its repair policies is limited to the final signal.

\textbf{Impacts of Feedback.} \quad We remove feedback from the reward function during process supervision, which prevents the language model from using reinforcement learning for policy adjustment. When given a repair sequence, we force the language model to use the subsequent repair step as a supervision signal for learning.

\textbf{Impacts of Reward Function.} \quad We introduce two variants to verify the effectiveness of pairwise ranking in reward modeling. We use two ranking methods: point-wise, and list-wise.
Point-wise ranking scores individual program states without considering their order relationship;
while list-wise ranking compares multiple programs together (beyond pairs) to determine their order relationship.

The results of the experiment are shown in Table~\ref{tab:ablation}. 
We can draw the following conclusions from the results: (1) Regardless of the supervision method used, introducing feedback to adjust LM's repair policies is necessary. 
Performance may significantly decrease due to additional noise if feedback is lacking in the process-based supervision method.
(2) Pair-wise ranking proves more effective as a reward model compared to point-wise or list-wise ranking.
\begin{figure}[t]
    \centering
    \includegraphics[width=0.48\textwidth]{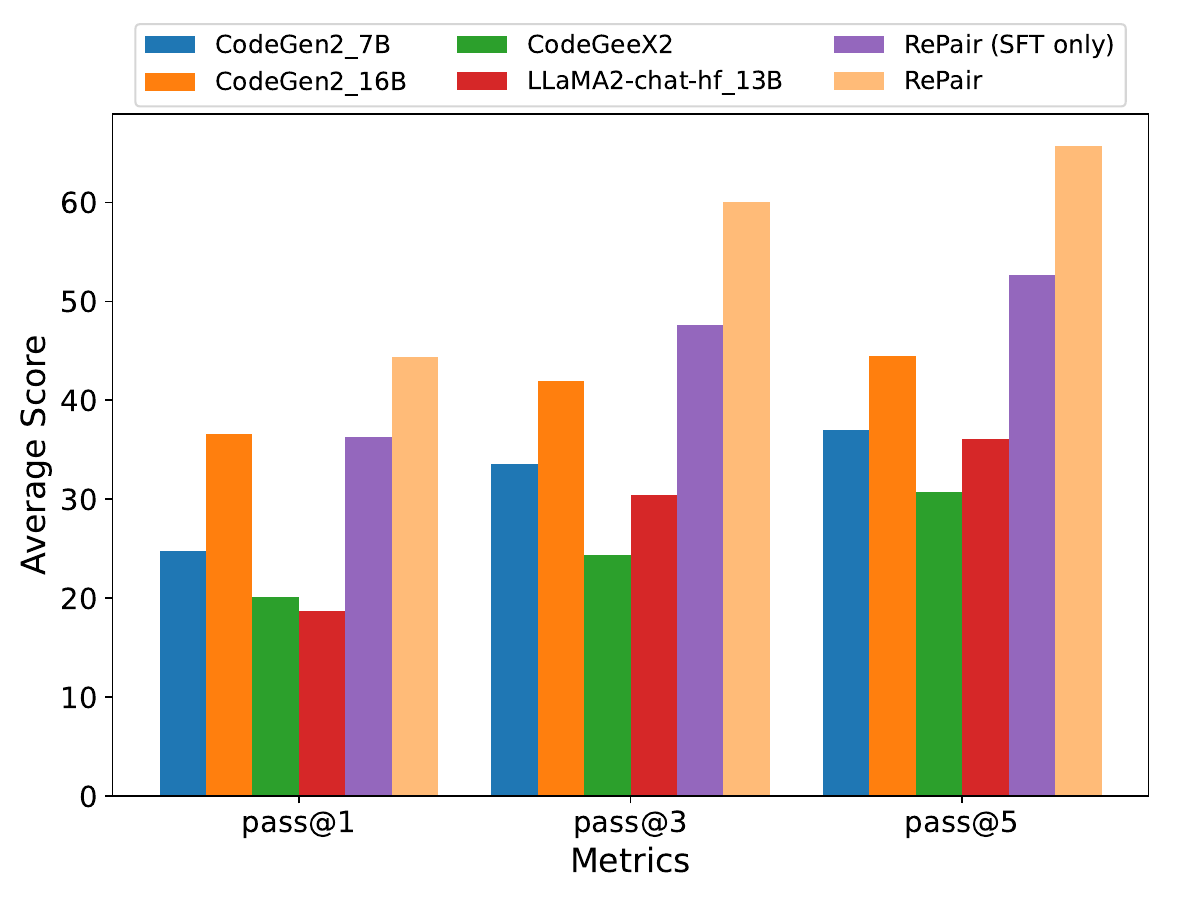}
    \caption{Performance comparison on other fine-tuned open-sourced model.}
    \label{sft-open}
\end{figure}
Using point-wise ranking solely requires the model to score each program based on compiler and test case, without comprehending the program's preferences and modification direction.
Conversely, providing $K$ rankings simultaneously in a list-wise task can be overly complicated and potentially hinder performance.
\subsection{Process-based Feedback Can Bring Performance Benefits}
Someone might question the superiority of RePair over current open-source models due to our use of SFT on the training set, while other models employed a few-shot setting. It is important to clarify that SFT was utilized to enhance the model's comprehension of the repair task, rather than being the primary factor for performance improvement. To substantiate this, we also fine-tune several open-source models with similar parameter counts using the same dataset. The results, presented in Figure~\ref{sft-open}, demonstrate a performance enhancement following SFT. However, this improvement is largely due to the model's increased understanding of the task and the standardization of output formats. Furthermore, when combined with SFT, process-based feedback method significantly amplifies the model's potential, leading to even greater performance gains.
\subsection{Smaller Models May Struggle to Receive Effective Explicit Feedback}
\begin{figure}[t]
    \centering
    \includegraphics[width=0.48\textwidth]{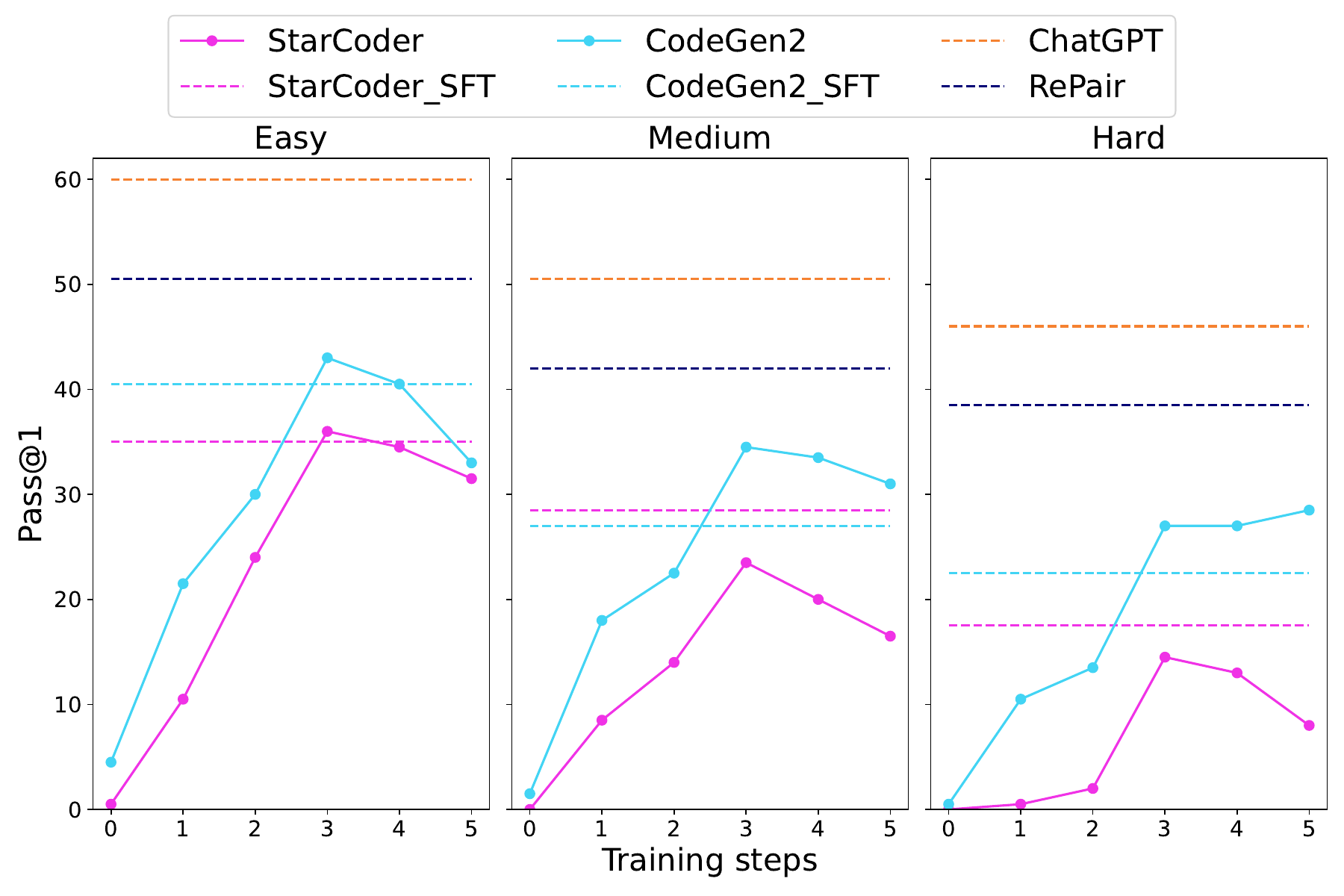}
    \caption{Performance of explicit prompts at different training steps. At three difficulty levels, the best performance was achieved at three training steps. ChatGPT can effectively understand feedback from the compiler and test cases. For small-scale models, explicit prompts are still difficult to understand.}
    \label{prompt}
\end{figure}
As mentioned in the abstract and introduction, understanding explicit prompts for small-scale LMs remains challenging. We directly interact with LMs using the prompt template provided in Appendix~\ref{sec:template}. We fine-tune two open-source models, StarCoder and CodeGen2, within 5 steps and present the experimental results in Figure~\ref{prompt}. The experimental results indicate that: (1) The performance of the open-source model peaked after three steps of fine-tuning. (2) ChatGPT's performance was further enhanced after receiving feedback from compilers and test cases, demonstrating better results than one-step repair. (3) Small-scale models struggle to comprehend feedback information through explicit prompts.
\subsection{Qualitative Results}
\begin{figure}[t]
    \centering
    \includegraphics[width=0.96\columnwidth]{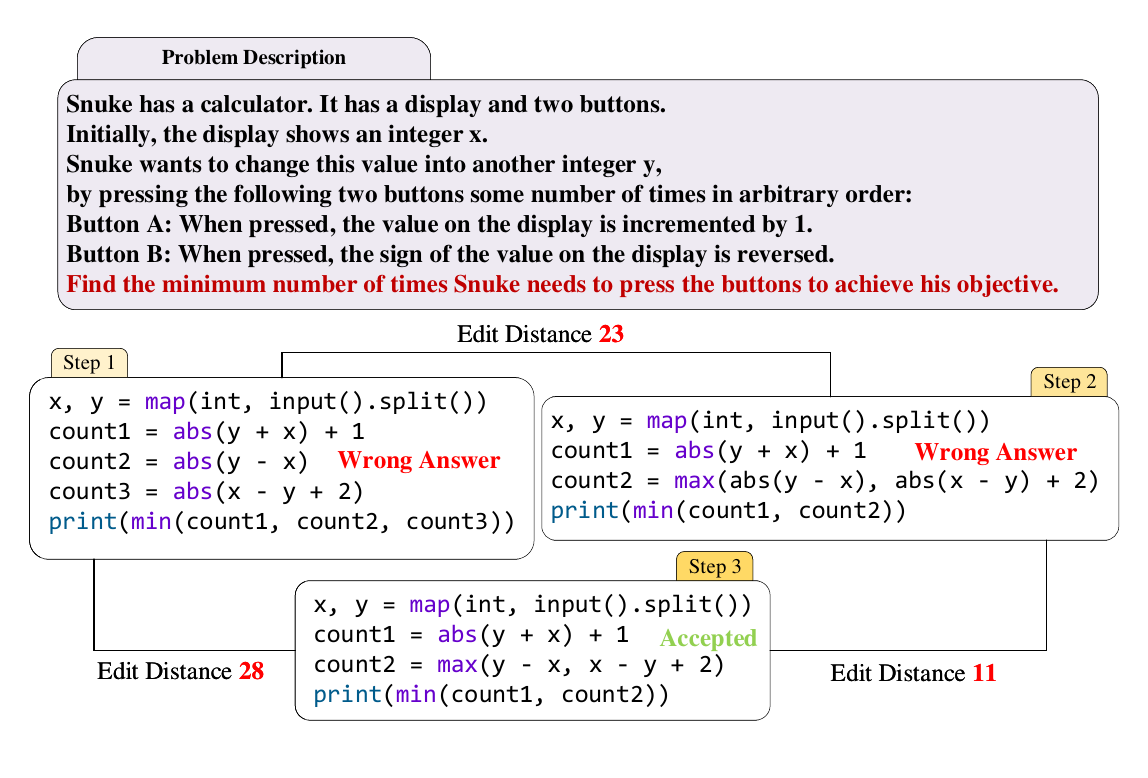}
    \caption{An example of repair process for ``Count the number of integer transitions''.}
    \label{case1}
\end{figure}
As shown in Figure~\ref{case1}, our model through step-by-step modification, have effectively simplified the complexity of each repair step and achieved satisfactory ``Accepted'' results.
For more examples, please refer to the Appendix~\ref{sec:cases}.
\section{Related Works}
\subsection{Large Language Models}
The recent years have witnessed a substantial evolution in the field of large language models.
language modeling is the prediction of the probability distribution of the next token given a context. The scaling law makes it possible to enhance performance by increasing the size of the model and data~\cite{kaplan2020scaling}.
The emergence of large-scale language models with over 100B parameters has consistently resulted in record-breaking performance~\cite{srivastava2023imitation, hendrycks2021measuring, chen2021evaluating}.
Larger models begin to solve problems that smaller models cannot handle, a phenomenon known as \textit{emergence}.
Typical phenomena of emergence include In Context Learning (ICL) and Chains of Thought (CoT)~\cite{brown2020language, wei2022chain}. 
The core of ICL is to draw knowledge from analogies.
After providing demonstrations, the query problem is presented alongside prompts to form the input.
A CoT is a prompt formed through a series of logical thinking steps, guiding the LLMs to answer in a thoughtful manner. This method can significantly improve performance on reasoning tasks~\cite{liu2023learning}.
Our work is dedicated to using small-scaled language models to enhance program repair performance.

\subsection{Outcome- and Process- based Approaches}
The outcome-based approaches supervise the final results, while the process-based approach focuses on each step leading to the result. Both have their strengths and weaknesses: outcome-based requires fewer supervision signals, but process-based approaches align more with human thinking patterns. In \cite{uesato2022solving}, researchers conducted experiments on math dataset and found that both approaches produced similar error rates, but due to cost considerations, the process-based method performed better. However, recent research~\cite{lightman2023lets} has proposed a different view; after building larger datasets and annotating more processes, the process-based method showed clear advantages. Process-based can help machine learning systems accurately understand humans' way of thinking and regulate potential hallucination.

\subsection{Learning from Feedback}
Through learning from feedback, language models can effectively receive signals beyond labels~\cite{gou2024critic}.
Some NLP tasks rely on automated evaluation metrics such as BLEU, ROUGE to get feedback from ground truth.
However, this kind of automatic feedback that solely depends on exact matching is not an effective assessment criterion~\cite{Colombo2022infolm}. 
Therefore, researchers are attempting to train language models using human preferences as supervisory signals.
In some open text generation tasks, learning from human feedback has significantly improved performance~\cite{jaques2019way, bahdanau2016actor, lawrence2018improving, zhou2020learning, stiennon2022learning}.
Following this line, \cite{ouyang2022training} applies human feedback to GPT-3 to achieve alignment with humans, thereby mitigating the issue of generating untruthful or toxic text.
Our work focuses on obtaining feedback from the compiler as a \textit{virtual tool}, integrating signals from the tools into the training process of language models.

\section{Conclusion}
In this work, we proposed an automated program repair method that utilizes process-based feedback.
This approach is inspired by strategies used in programming competitions and incorporates process supervision into the repair process.
Due to the unavailability of suitable datasets, we first created a multi-step repair dataset called CodeNet4Repair.
This dataset not only includes the initial faulty programs and the final accepted ones but also records the intermediate repair process.
On this basis, we implemented supervision over the process with feedback. Specifically, we fine-tuned a pre-trained model StarcoderBase through to make it understand program repair tasks.
Then, we empirically defined a partial order relationship for program states and trained reward models using pairwise ranking. This reward model would serve as a critic, assessing each attempt by the LM to polish the program.
Meanwhile, the LM plays the role of an actor, constantly adjusting its policies for repairing the program based on received rewards.
During the inference stage, LM iteratively repaired the program until the exit conditions are met.
The experimental results indicated that the process-based feedback method can demonstrate excellent performance.
With smaller model sizes, \textbf{RePair} still can rival or even surpass closed-source commercial large models.
We believe that process-based feedback is currently underexplored, and will continue to explore more universal methods.

\section{Limitation}
We contributed a dataset called CodeNet4Repair, which focuses on program repair in real competition scenarios. We acknowledge that CodeNet4Repair still has limitations in evaluating model repair capabilities. Due to time and resource constraints, we were unable to collect all repair processes across different languages and software engineering domains. But there are significant similarities between competition and engineering scenarios.
This implies that our model can achieve a level of usability for engineering project purposes, and our dataset can partially evaluate its repair performance in an engineering context.

\section{Acknowledgement}
This research was partially supported by grants from the National Natural Science Foundation of China (No. 62106244), the Anhui Provincial Natural Science Foundation (No. 2308085QF229), the Fundamental Research Funds for the Central Universities (No. WK2150110034), and the CIPSC-SMP-Zhipu.AI Large Model Cross-Disciplinary Fund.
\bibliography{anthology,custom}
\newpage
\appendix
\section{Appendix}
\label{sec:appendix}
\subsection{Prompts Used in Experiments}
\begin{figure}[htbp]
    \centering
    \includegraphics[width=0.48\textwidth]{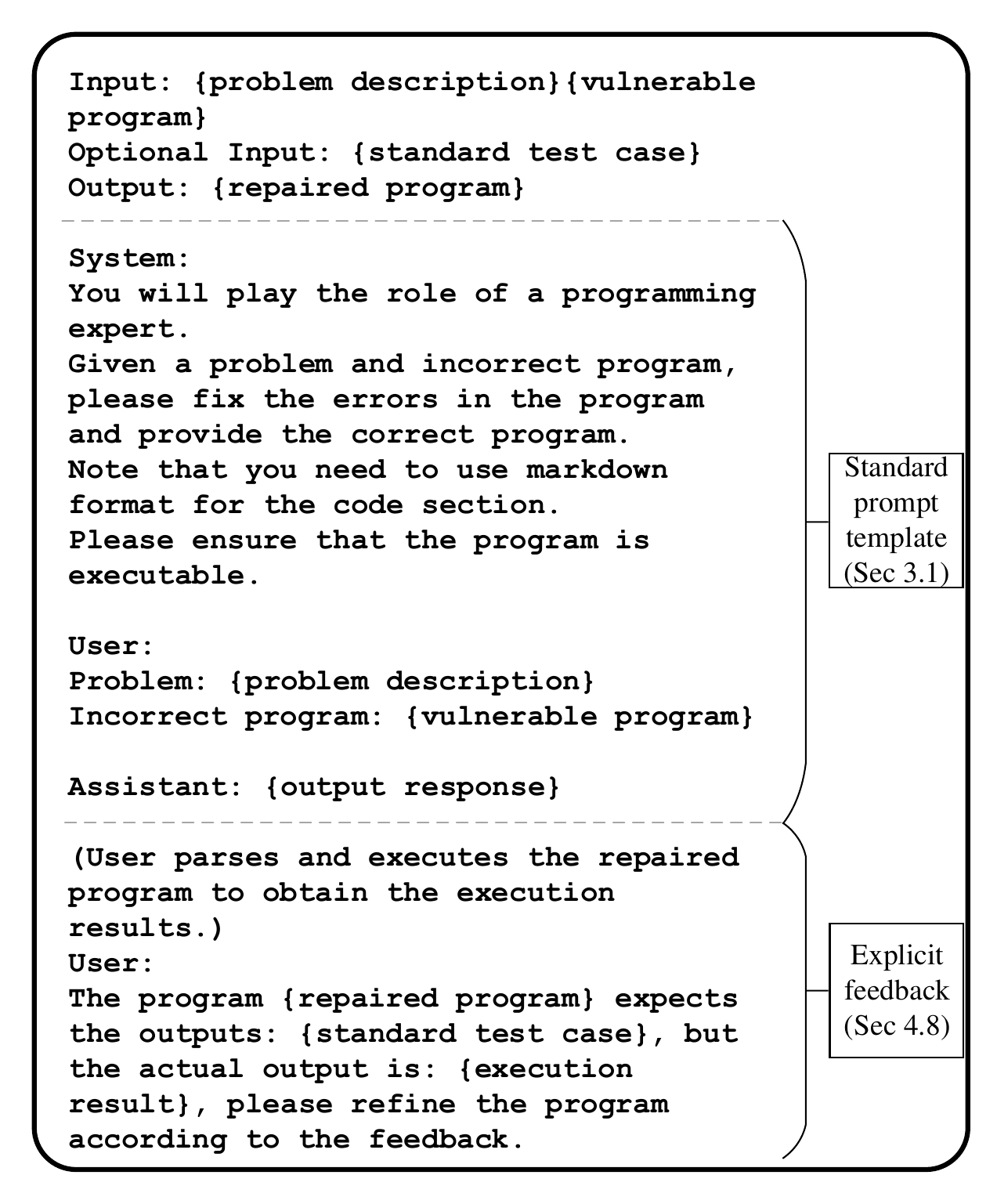}
    \caption{Prompt Template for Program Repair.}
    \label{prompt_template}
\end{figure}

We performed supervised fine-tuning and explicit feedback experiments following the prompt template shown in Figure~\ref{prompt_template}. For explicit feedback, we also input optional standard test cases.
\label{sec:template}
\subsection{Ethics Statement}
As creators and users of LLMs, we recognize the profound ethical responsibilities that come with their development and deployment.
We will strive to avoid creating or perpetuating harm, including the reinforcement of bias, misinformation, and harmful stereotypes.
We advocate for the use of open-source LLMs in potential ethical risk research.
We believe that a public LLM can reveal its internal state, thereby avoiding possible risks.
\subsection{Detailed Case Study}
\label{sec:cases}
In the next page, we provide six examples of repairs.
Through the examples shown, we found that the edit distance for single-step modifications (directly from Step 1 to Step 3) is very high, resulting in tasks being too complex (e.g., having both compilation errors and semantic errors) to be completed in one step.
However, process-based modifications can decompose complex tasks into simple ones and solve each problem gradually, thereby producing good repair effects.
\begin{figure*}[t]
    \centering 
\includegraphics[width=0.9\textwidth]{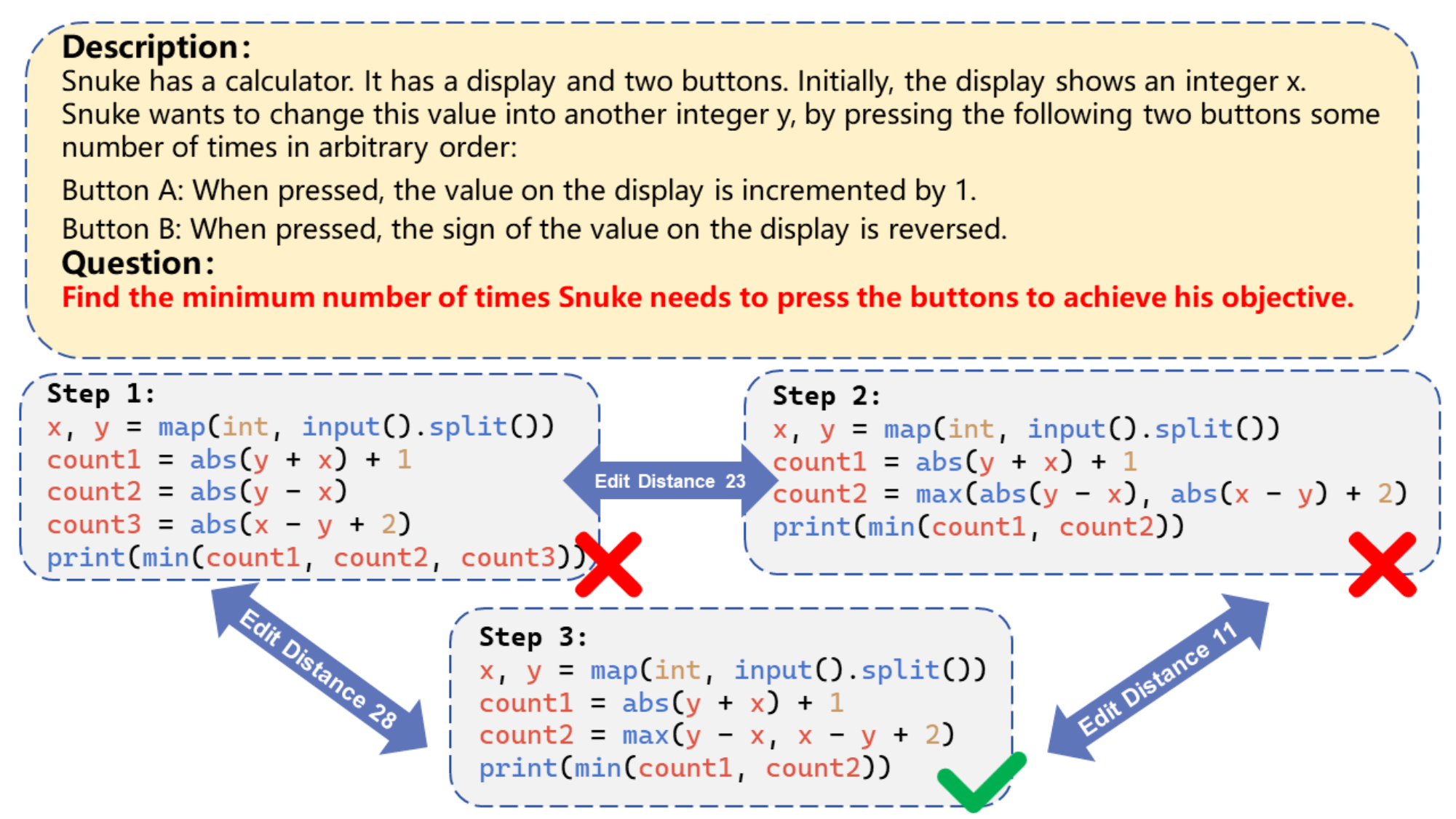}
\caption{Case 1, an example of repair process for ``Count the number of integer transition''.}
\end{figure*}
\begin{figure*}[t]
    \centering 
\includegraphics[width=0.9\textwidth]{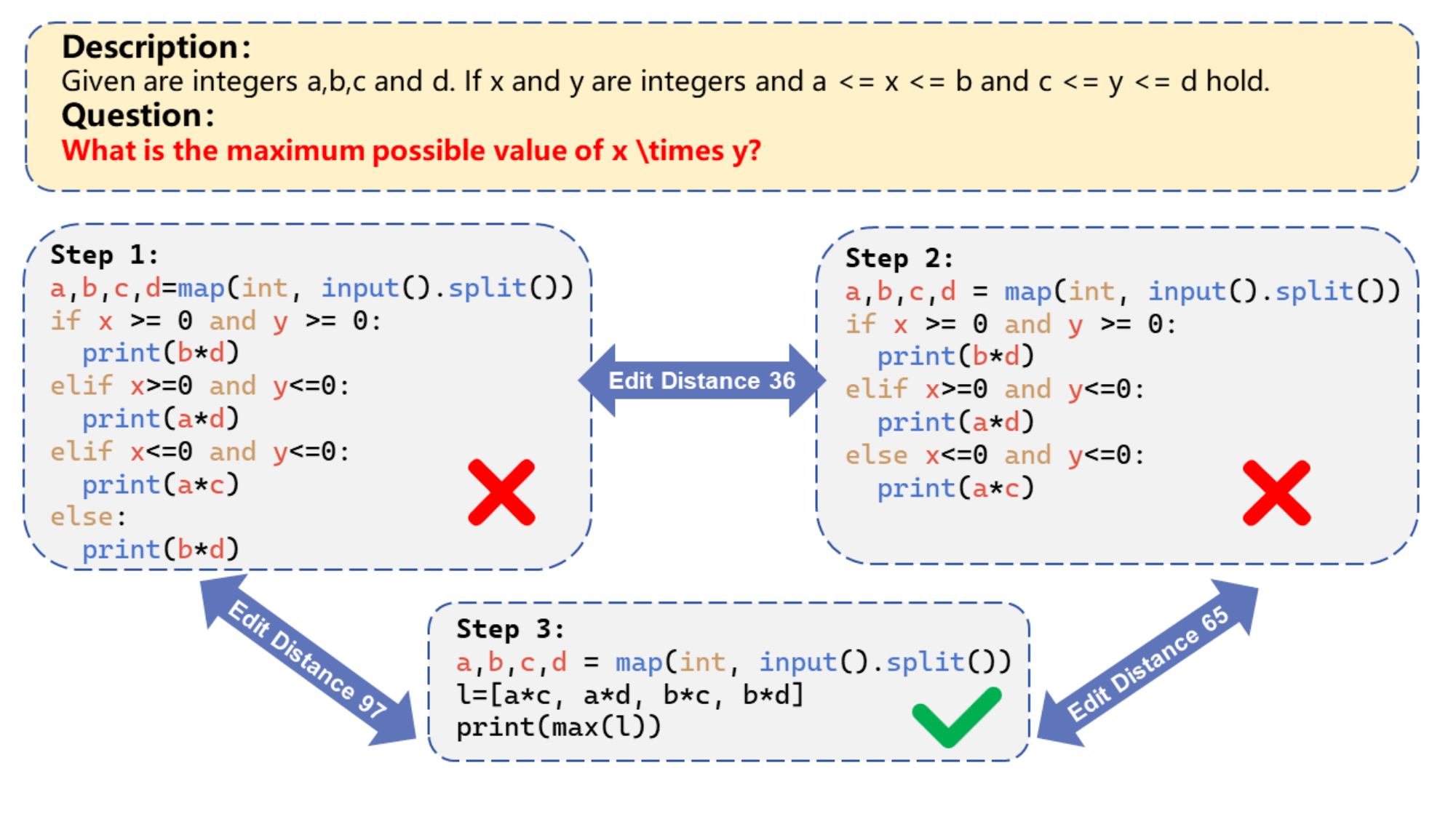}
\caption{Case 2, an example of repair process for ``Maximum possible value of x $\times$ y''.}
\end{figure*}
\begin{figure*}[t]
    \centering 
\includegraphics[width=0.9\textwidth]{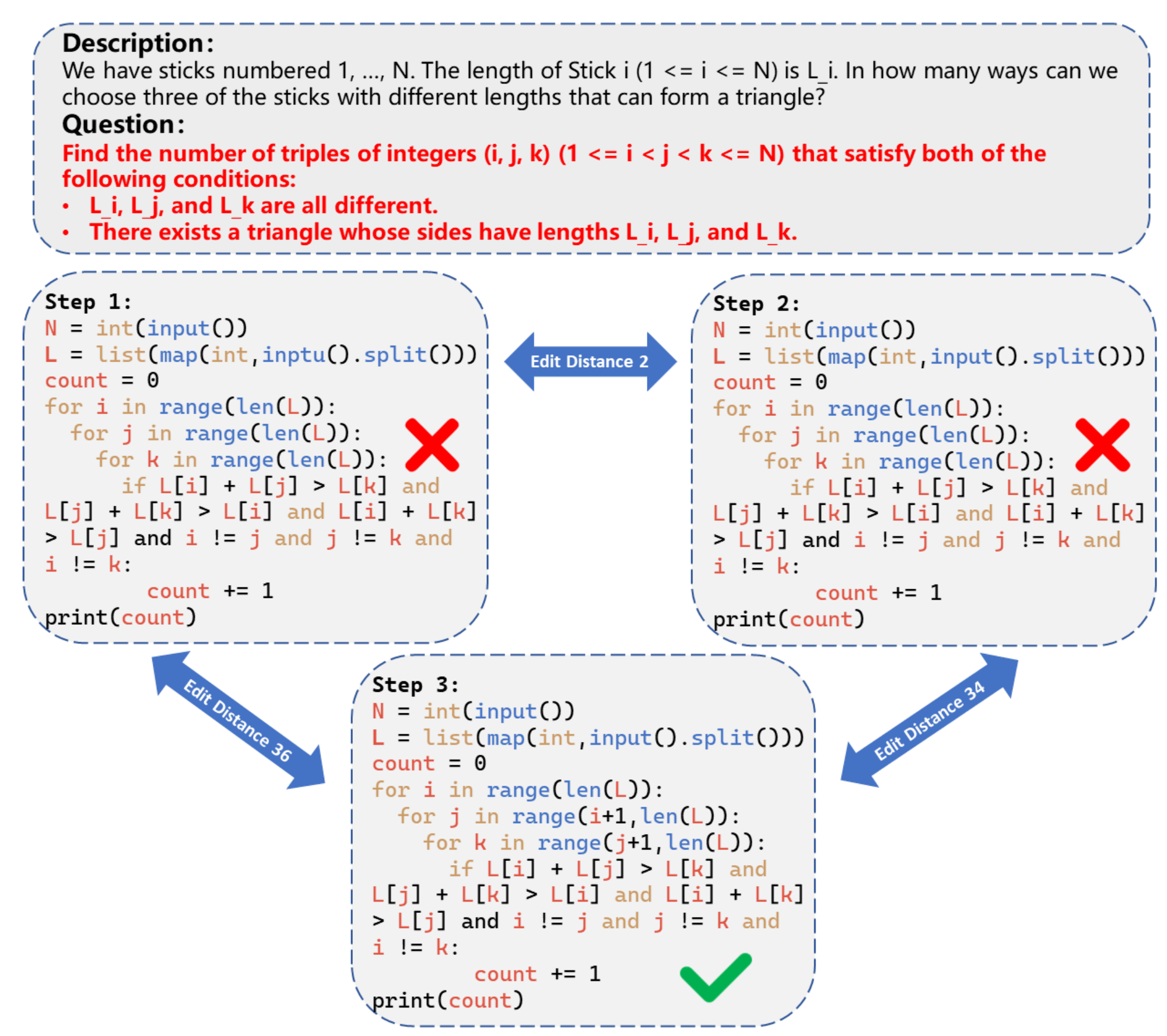}
\caption{Case 3, an example of repair process for ``The number of combinations of sticks that can form a triangle''.}
\end{figure*}
\begin{figure*}[t]
    \centering 
\includegraphics[width=0.9\textwidth]{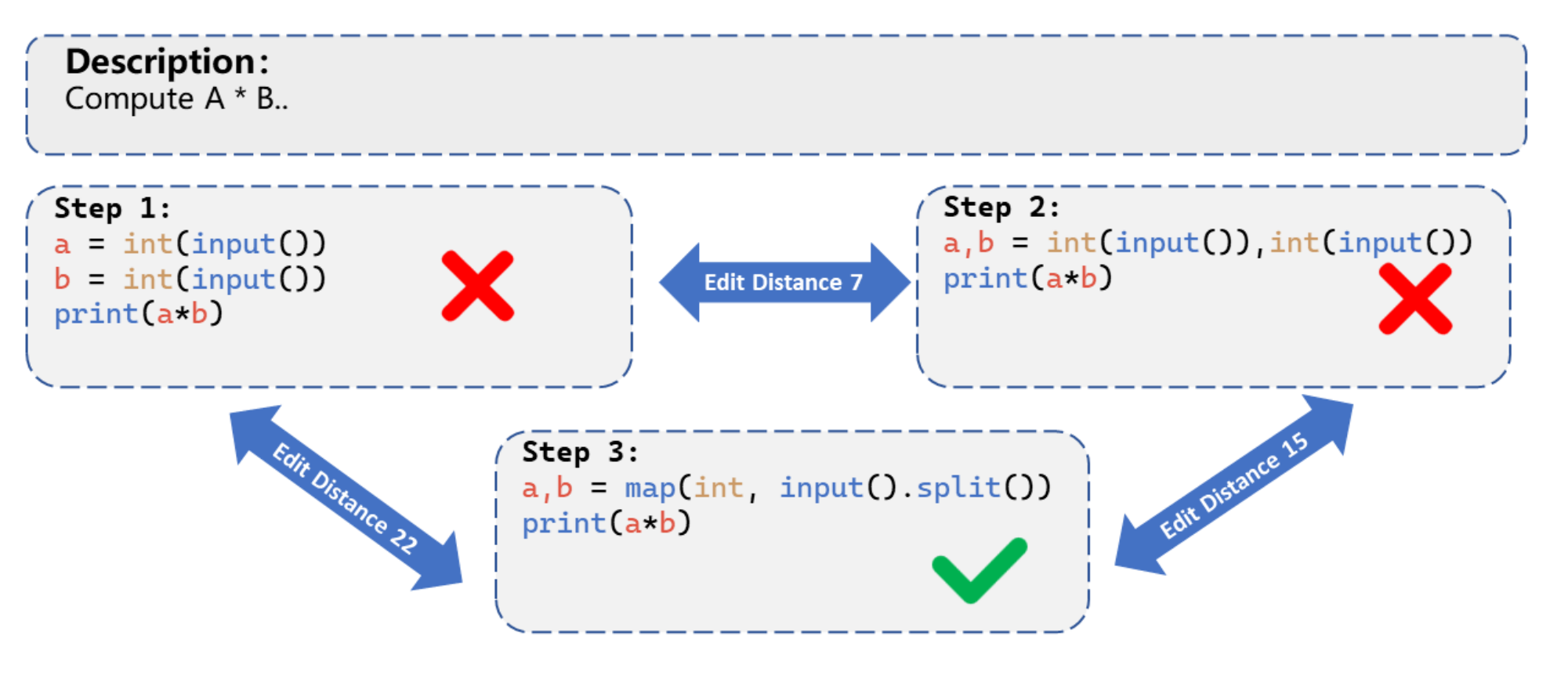}
\caption{Case 4, an example of repair process for ``A $\times$ B''.}
\end{figure*}
\begin{figure*}[t]
    \centering 
\includegraphics[width=0.9\textwidth]{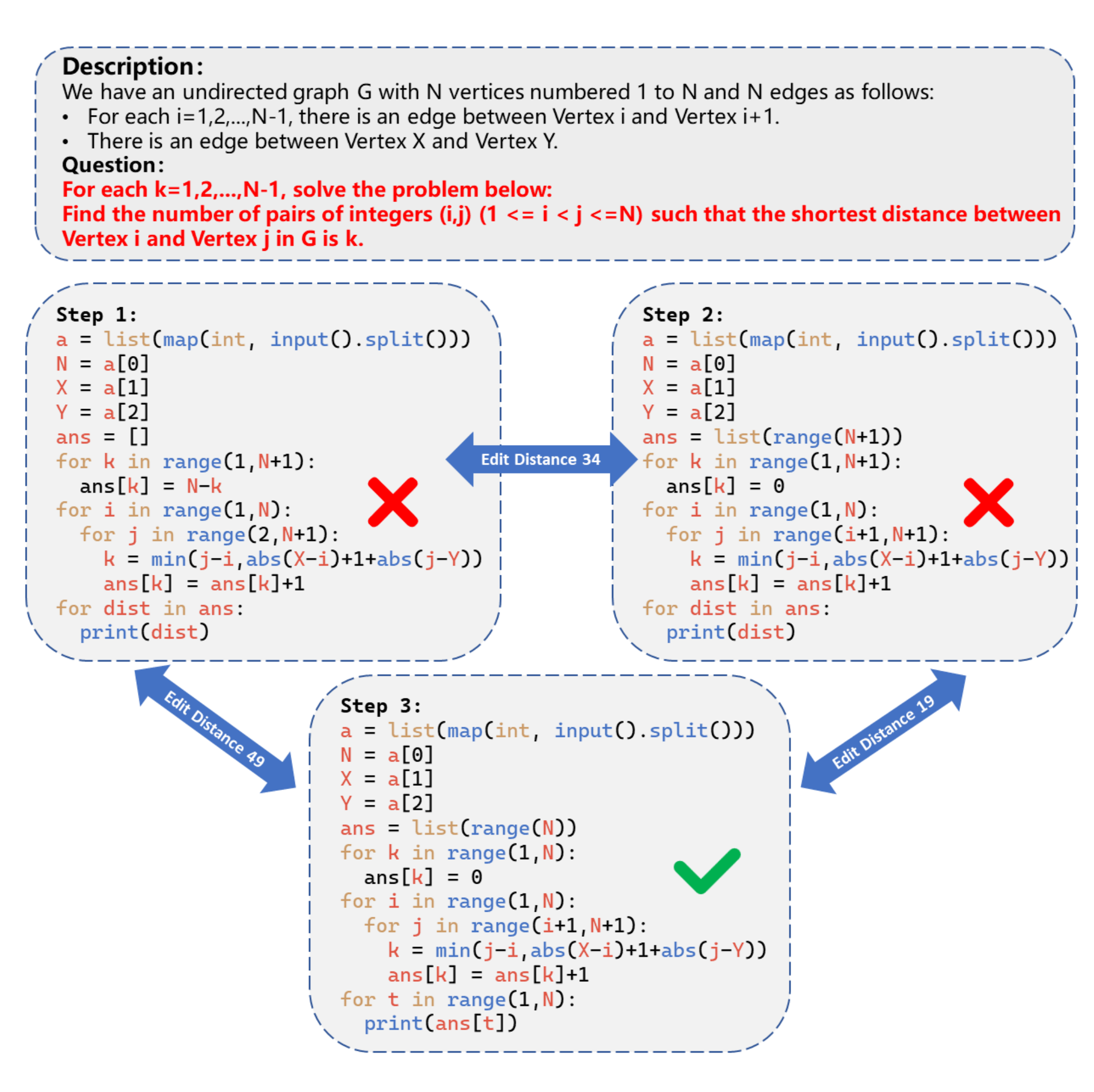}
\caption{Case 5, an example of repair process for ``Distance calculation in an undirected graph''.}
\end{figure*}
\begin{figure*}[t]
    \centering 
\includegraphics[width=0.9\textwidth]{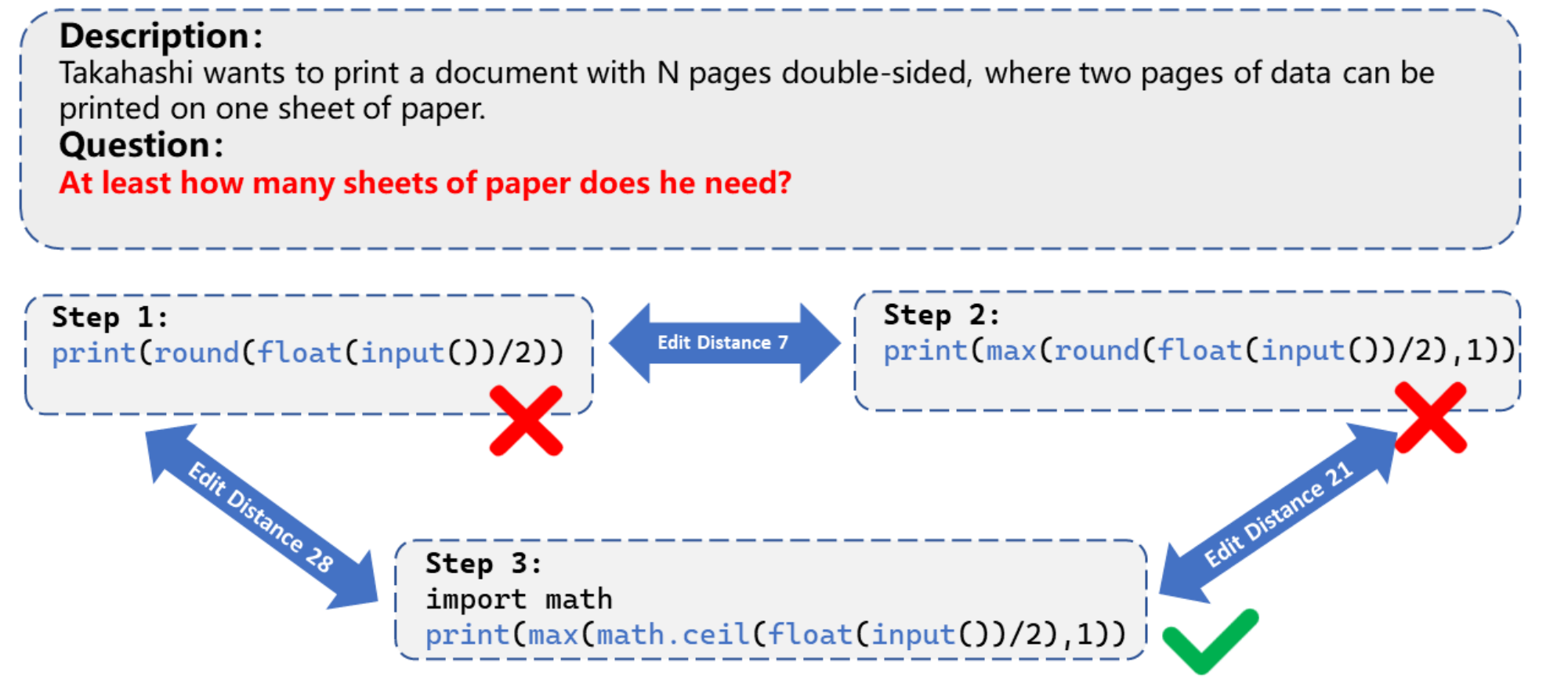}
\caption{Case 6, an example of repair process for ``Calculate the minimum number of sheets of paper required''.}
\end{figure*}
\end{document}